\documentclass[superscriptaddress,showpacs, showkeys, preprint, floatfix]{revtex4}
\usepackage{graphicx}
\begin{document}
\title{$\pi^\pm p$ differential cross sections at low energies}
\date{\today}
\author{H. Denz}
\email{denz@pit.physik.uni-tuebingen.de}
\affiliation{Physikalisches Institut, Universit\"at T\"ubingen, 72076 T\"ubingen, Germany}
\author{P. Amaudruz}
\affiliation{TRIUMF, Vancouver, British Columbia, Canada V6T 2A3}
\author{J.T. Brack}
\affiliation{University of Colorado, Boulder, Colorado 80309-0446, U.S.A.}
\author{J. Breitschopf}
\affiliation{Physikalisches Institut, Universit\"at T\"ubingen, 72076 T\"ubingen, Germany}
\author{P. Camerini}
\affiliation{Dipartimento di Fisica dell'Universita' di Trieste, I-34127 Trieste, Italy}
\affiliation{Istituto Nazionale di Fisica Nucleare, I-34127 Trieste, Italy}
\author{J.L. Clark}
\affiliation{School of Physics, University of Melbourne, Parkville, Victoria 3052, Australia}
\author{H. Clement}
\affiliation{Physikalisches Institut, Universit\"at T\"ubingen, 72076 T\"ubingen, Germany}
\author{L. Felawka}
\affiliation{TRIUMF, Vancouver, British Columbia, Canada V6T 2A3}
\author{E. Fragiacomo}
\affiliation{Istituto Nazionale di Fisica Nucleare, I-34127 Trieste, Italy}
\author{E.F. Gibson}
\affiliation{California State University, Sacramento, California 95819, U.S.A.}
\author{N. Grion}
\affiliation{Istituto Nazionale di Fisica Nucleare, I-34127 Trieste, Italy}
\author{G.J. Hofman}
\affiliation{TRIUMF, Vancouver, British Columbia, Canada V6T 2A3}
\affiliation{University of Regina, Regina, Saskatchewan, Canada S4S 0A2}
\author{B. Jamieson}
\affiliation{TRIUMF, Vancouver, British Columbia, Canada V6T 2A3}
\author{E.L. Mathie}
\affiliation{University of Regina, Regina, Saskatchewan, Canada S4S 0A2}
\author{R. Meier}
\affiliation{Physikalisches Institut, Universit\"at T\"ubingen, 72076 T\"ubingen, Germany}
\author{G. Moloney}
\affiliation{School of Physics, University of Melbourne, Parkville, Victoria 3052, Australia}
\author{D. Ottewell}
\affiliation{TRIUMF, Vancouver, British Columbia, Canada V6T 2A3}
\author{O. Patarakin}
\affiliation{Kurchatov Institute, Moscow, Russia}
\author{J.D. Patterson}
\affiliation{University of Colorado, Boulder, Colorado 80309-0446, U.S.A.}
\author{M.M. Pavan}
\affiliation{TRIUMF, Vancouver, British Columbia, Canada V6T 2A3}
\author{S. Piano}
\affiliation{Dipartimento di Fisica dell'Universita' di Trieste, I-34127 Trieste, Italy}
\affiliation{Istituto Nazionale di Fisica Nucleare, I-34127 Trieste, Italy}
\author{K. Raywood}
\affiliation{TRIUMF, Vancouver, British Columbia, Canada V6T 2A3}
\author{R.A. Ristinen}
\affiliation{University of Colorado, Boulder, Colorado 80309-0446, U.S.A.}
\author{R. Rui}
\affiliation{Dipartimento di Fisica dell'Universita' di Trieste, I-34127 Trieste, Italy}
\affiliation{Istituto Nazionale di Fisica Nucleare, I-34127 Trieste, Italy}
\author{M.E. Sevior}
\affiliation{School of Physics, University of Melbourne, Parkville, Victoria 3052, Australia}
\author{G.R. Smith}
\affiliation{TRIUMF, Vancouver, British Columbia, Canada V6T 2A3}
\affiliation{Jefferson Lab, Newport News, VA 23006 U.S.A.}
\author{J. Stahov}
\affiliation{University Tuzla, Faculty of Science, 35000 Tuzla, Bosnia and Herzegovina}
\author{R. Tacik}
\affiliation{University of Regina, Regina, Saskatchewan, Canada S4S 0A2}
\author{G.J. Wagner}
\affiliation{Physikalisches Institut, Universit\"at T\"ubingen, 72076 T\"ubingen, Germany}
\author{F. von Wrochem}
\affiliation{Physikalisches Institut, Universit\"at T\"ubingen, 72076 T\"ubingen, Germany}
\author{D.M. Yeomans}
\affiliation{University of Regina, Regina, Saskatchewan, Canada S4S 0A2}
\begin{abstract}
Differential cross sections 
for $\pi^-p$ and $\pi^+p$ elastic scattering 
were measured at five energies
between 19.9 and 43.3 MeV. The use of the CHAOS magnetic spectrometer
at TRIUMF, supplemented by a range telescope for muon background
suppression, provided simultaneous coverage of a large part of  the full 
angular
range, thus allowing very precise relative cross section measurements. 
The absolute normalisation was determined with a typical accuracy of 5 \%.
This was verified in a simultaneous measurement of muon 
proton elastic scattering. The measured cross sections show some deviations
from phase shift analysis predictions, in particular
at large angles and low energies. From the new data we determine the real part of the isospin forward scattering amplitude.
\end{abstract}
\maketitle
Pion-nucleon scattering at low energies allows the study of
non-perturbative aspects of QCD on one of the simplest hadronic
systems. The prime example is the determination of the $\pi$N-sigma
term which is a measure of the explicit breaking of chiral symmetry
through non-vanishing quark masses \cite{Gas91}. However, the \
reported values
range from the canonical 64 MeV \cite{Koch} to about 80 MeV \cite{Pavan} and 
there is a
longstanding dispute whether this scatter is due to the
method of extraction or due to the data base used (or both). A solution of the puzzle
is highly desirable, all the more since, with the conventional understanding
 \cite{Gas91}, 
values around 80 MeV would imply
a strange sea quark content of the nucleon which is at variance with our
current understanding of its structure.

Independent of theoretical considerations the data base at pion kinetic energies 
below 50
MeV is scarce and, where existing, sometimes contradictory. Low energy data are
of considerable importance since the determination of the sigma term
requires an extrapolation of the scattering amplitudes to the
unphysical Cheng-Dashen point \cite{CD71} below the $\pi$N threshold. 
While observables at the threshold (i.e. scattering lengths) are 
being determined 
by precision measurements of pionic hydrogen \cite{Schroeder}, the
energy dependence of the phase shifts, which is required for the
extrapolation, remains largely uncertain. This is due to the experimental
difficulties inherent in $\pi p$ scattering experiments at low
energies, where the pion decay lengths are small and the
resulting muon background presents a severe problem.

\begin{figure}
\includegraphics[width=\columnwidth]{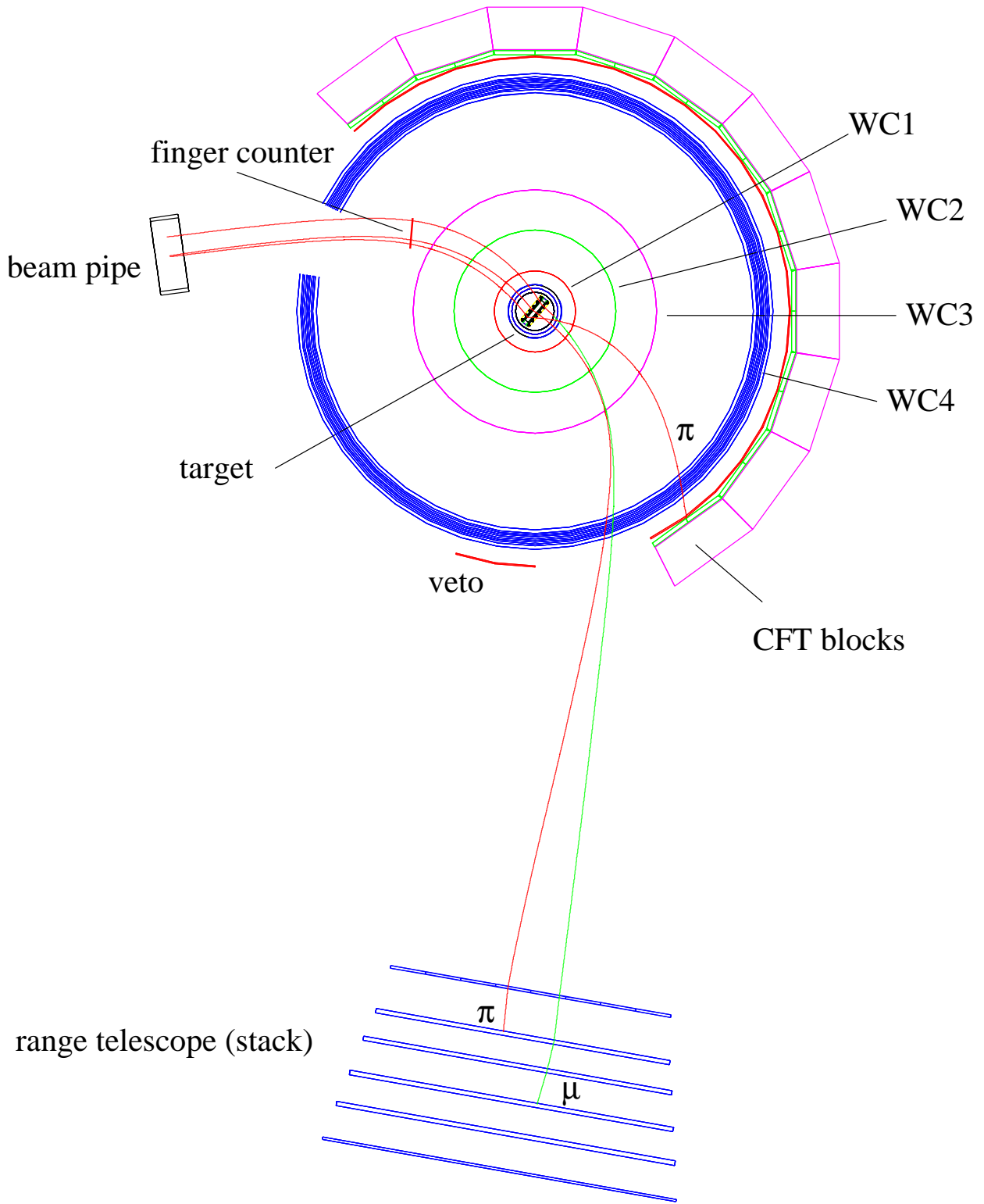}
\caption{Experimental setup: The magnetic field of CHAOS is oriented 
perpendicular to the plane of the figure. The fourth wire chamber 
(WC4) is surrounded by plastic scintillator and lead glass Cerenkov counters (CFTs). The range telescope consists of 6 layers of plastic scintillators and covers the forward scattering angles. Three typical events are plotted,
a scattered pion detected by a CFT block, a scattered pion detected by the
range telescope and a pion decaying into a muon in the target region which
is then detected by the range telescope.
}
\label{chaos}
\end{figure}

The present experiment exploits the properties of the CHAOS
spectrometer \cite{Smi95} at the M13 low-energy pion channel of TRIUMF. The
spectrometer is well suited for low energy pion scattering measurements since 
it has a
compact design and allows a simultaneous measurement of almost the full angular
distribution. Briefly, CHAOS (see
fig. \ref{chaos}) is a magnetic spectrometer with a $2\pi$  acceptance in the
reaction plane and a $\pm 7 ^\circ$ acceptance out-of-plane. It consists of
an axially symmetric dipole magnet with a pole diameter of 96 cm. The
target in its center is surrounded by four concentric rings of wire
chambers. This tracking region is surrounded by fast trigger
counters consisting of plastic scintillation counters and lead glass Cerenkov 
blocks. 
For this experiment (see fig. \ref{chaos}) blocks at forward 
scattering angles were
removed and replaced by a range telescope \cite{Fra00}. Information from 
this telescope was interpreted using a
software neural network to discriminate scattered pions from
the huge background of decay muons.

The target consisted of 80 cm$^3$ of liquid hydrogen contained in a cell with
flat rectangular Mylar windows 125 $\mu$m thick and 1.25 cm apart.
It was surrounded by an outer cell filled by hydrogen gas of the same
pressure to ensure flat target windows. Data were taken with and
without liquid hydrogen for background subtraction.

The experiment used positively and negatively charged pions with
energies of $19.9\pm0.3$, $25.8\pm0.3$, $32.0\pm0.3$, $37.1\pm0.4$, 
and $43.3\pm0.4$ 
MeV from the M13 channel of
TRIUMF. 
The uncertainties follow from a time of flight calibration of the pion channel 
using the method described in \cite{Pav01c}. 
Muons and electrons from the production target were
discriminated from pions by their time of flight as taken from the cyclotron RF
pulse and a time signal derived from a thin ``finger'' scintillation
counter (see fig. \ref{chaos}) at the entrance of CHAOS. 

The data were taken with a two-stage trigger system.
The first level
trigger required a hit in the finger counter with
the correct time of flight for pions or muons through the channel,  
no veto in any of the veto counters, and at least one
hit in the first layer of the range telescope or the CFT blocks. The
second level trigger rejected events with hit patterns typical for  
unscattered beam particles using information from the inner two wire chambers.

Incoming and outgoing particles were detected in the wire
chambers. Momenta, vertices and
scattering angles were reconstructed from the hits in the wire chambers.
It is noteworthy that the out-of-plane-component of the scattering
angle was also determined using cathode strips and resistive wires. 
Only events fulfilling the
kinematics of elastic pion-proton scattering were
accepted. Furthermore, valid vertices were required to lie in the liquid
hydrogen region. 
A range telescope \cite{Fra00} was installed in order to mitigate the 
otherwise large forward-angle background of muons from pion decay in the
target region.
The range telescope consisted of 6 layers 
of plastic scintillator. The first layer was segmented into 8 paddles
allowing angle-dependent prescaling of events at forward angles. The hit of the
kinematically correct paddle was also checked to ensure the absence of pion 
decay.
Depending on the beam energy, suitable aluminum absorbers were inserted
for the most sensitive response. 
Neural network training runs were taken with pions and muons identified by time of flight
in the channel and directed directly onto the individual paddles.
After training, the neural network achieved a 98 \% 
efficiency in pion-muon discrimination using the $\Delta$E, range and 
time of flight information of the telescope.

The (energy-dependent) acceptance of the set-up was determined by 
GEANT3 \cite{geant} Monte-Carlo simulations. Special care was taken to ensure a
correct detector and target model including all materials. 
This is especially important for the lowest energies where
energy losses
play a significant role. For the backward angles at the lowest energy 
(19.9 MeV) 
the high sensitivity to the choice of material and geometry 
prevented a reliable determination of the acceptance. 
At all energies angular regions of rapidly changing acceptance near the 
support pillars and the border 
between CFT and range telescope were discarded.
Regions where the decay muon background was more than two orders of 
magnitude larger than
the pion rate had to be discarded.
The usual corrections for deadtime of the data acquisition system, chamber
efficiencies, pion decay and pion flux reduction due to hadronic
events were also applied.

In order to check the acceptance and the absolute normalisation
of cross sections by lepton scattering \cite{Barnett}, incident muons were 
selected by their time of flight 
in the channel. 
Muon-proton differential cross sections
were measured at forward angles (up to 25 degrees) where they are
sufficiently large. They were compared to
calculated electromagnetic cross sections taking into account the
proton charge
distributions \cite{yuri}. We observed good agreement of the relative angular
distributions. The average ratio of measured to calculated differential
cross sections agreed with unity within an error 
of $\pm 5$ \% which we take as the normalisation error of the pion
cross sections. Exceptions are the data at 43 MeV where the error is
larger ($\pm 7$ \%) for statistical reasons, and at 37 MeV where the measured
muon cross sections are consistently low by 8 \% leading to an asymmetric
estimated normalization error ($+5,-9$ \%). 
Details of the experiment and the cross sections in numerical form 
may be found in ref. \cite{Den04}.

\begin{figure}
\includegraphics[width=\columnwidth]{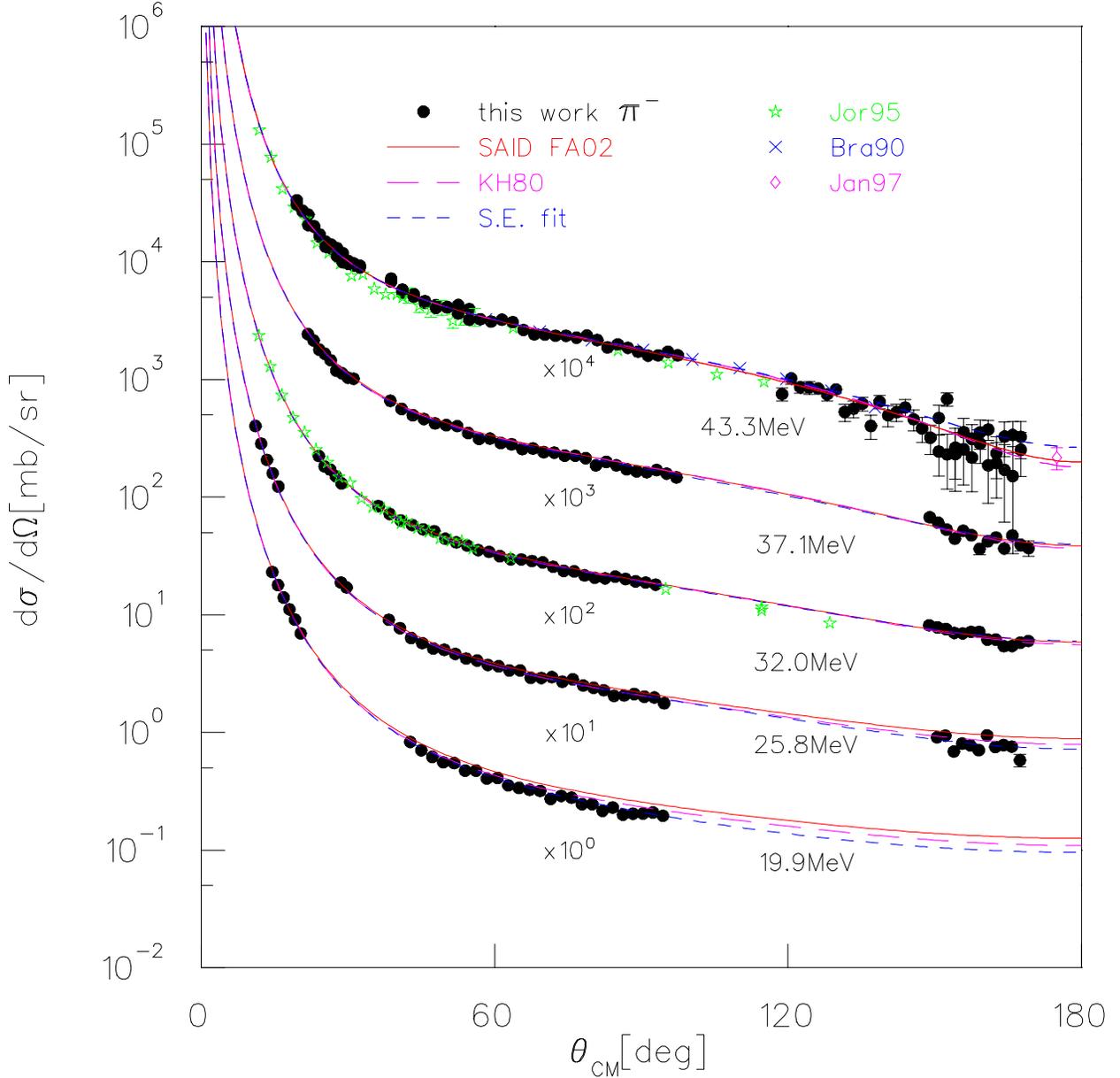}
\caption{Results of this experiment for $\pi^-p$ scattering
together with phase shift solutions and results from other experiments
at closeby energies. Bars denote 
statistical errors only. The absolute normalisation is uncertain by 5 to 9 \% 
(see text).}
\label{resminus}
\end{figure}

\begin{figure}
\includegraphics[width=\columnwidth]{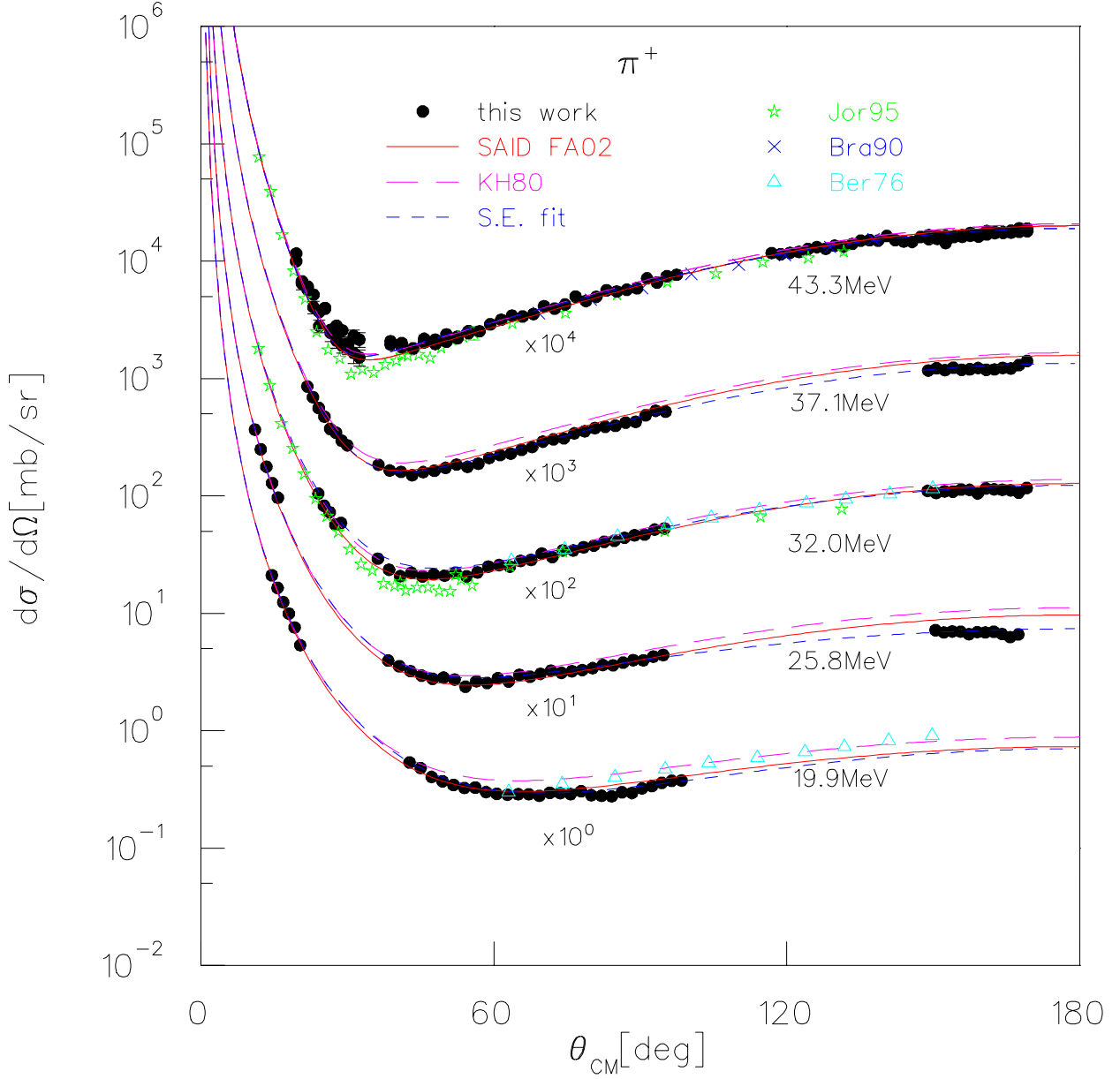}
\caption{As Fig. \ref{resminus}, but for $\pi^+p$ scattering.}
\label{resplus}
\end{figure}

The results of the present experiment are summarized in figs. \ref{resminus} 
and \ref{resplus} for $\pi^-p$ and $\pi^+p$ scattering, respectively. 
A complete PSA of the cross sections over the full energy range with
amplitudes constrained by analyticity and dispersion relations is
beyond the scope of this paper. Instead single-energy (SE) fits to our data 
were made. At each energy the phase shifts for S- and
P-waves were adjusted  simultaneously for $\pi^-p$ and $\pi^+p$ scattering and 
the phase shifts for D- and F-waves were taken from the KH84 \cite{Koc85} 
solution and kept fixed.  For comparison,
figs. \ref{resminus} and \ref{resplus} show the predictions from the SAID 
FA02 \cite{SAID}, KH80 \cite{KH80} phase
shift analysis (PSA) together with the SE fits. 
At first glance the data show an impressive overall
agreement with the predictions. A closer look, however,  reveals a 
general trend of
the PSA predictions to lie above the data  at low energies and
large angles.  The SE fits on the other hand are able to reproduce the data.

At energies below 30 MeV SAID tends to overestimate the  
$\pi^-p$ cross sections at
large scattering angles, where the KH80 solution and the SE fit 
give better descriptions. A comparison with
previous data shows that near 43 MeV our data and also the SAID
solution lie just in-between the angular distributions as measured by Brack
et al. \cite{Bra90} and by Joram et al. \cite{Jor95}, respectively. 
The suppression near 40
degrees observed in the latter work is not seen, whereas the 175 degree data
point by Janousch et al. \cite{Janousch} is confirmed. Near 32 MeV our cross 
sections agree
with the PSA predictions  whereas the results of Joram et al. \cite{Jor95} fall
somewhat low beyond 80 degrees.

The situation for $\pi^+p$ scattering is much more difficult.
Near 25 degrees at 43.3 MeV the CNI minimum  is substantially filled in, 
which is not seen in the three PSA results. 
At 19.9 MeV the CNI depression of the data is stronger than predicted
by KH80.
The simultaneously
taken $\mu^+p$ cross sections and also the $\pi^-p$ data  
do not show such an excess which suggests that it 
is not an artifact of the analysis.
At larger angles the agreement with the 45.0 MeV data by Brack et al. 
\cite{Bra90} is satisfactory.
The data of Joram et al. \cite{Jor95} near 45 MeV and 32 MeV exhibit an
even deeper minimum than the SAID fit and fall substantially below our data at
backward angles.
A general feature of our data
is that the slope of the angular distributions (relative to the SAID fits)
increases with decreasing energy. The discrepancy between our results
and those of Bertin et al. \cite{Ber76} at 20.8 MeV is obvious.
Clearly this data supports the previous criticism of the data sets \cite{Jor95}
and  \cite{Ber76} by Fettes and Matsinos \cite{Mat97}.

\begin{figure}
\includegraphics[width=\columnwidth]{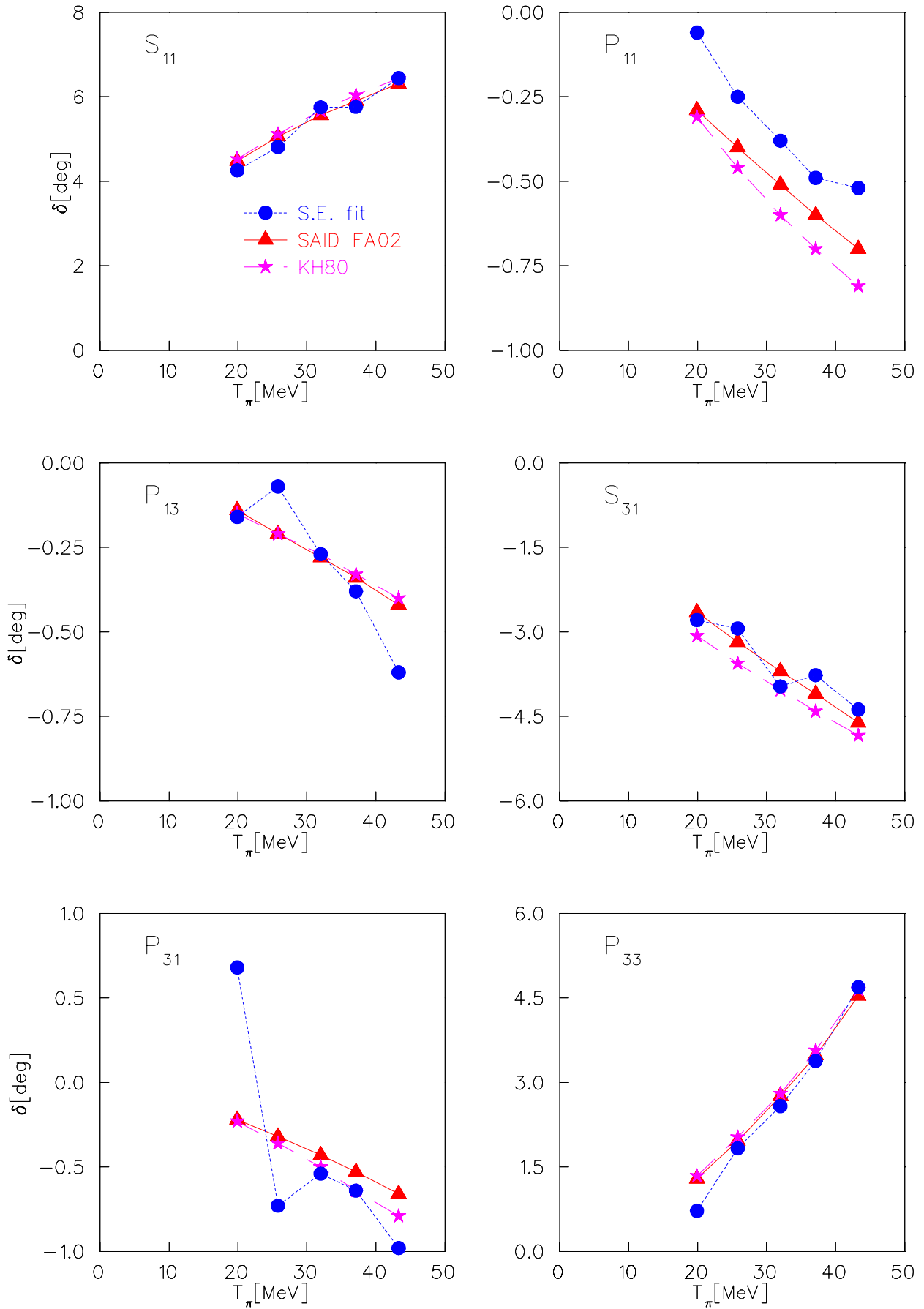}
\caption{Energy dependence of s- and p-wave phase shifts}
\label{phases}
\end{figure}

As shown in fig. \ref{phases}, the $S_{11}$, $S_{31}$ and $P_{33}$ 
phase shifts determined in the SE fits are
very close to the SAID FA02 and KH80 solutions, with the former 
being slightly favoured by the $S_{31}$ phases. 
This agreement is somewhat in contrast
to the findings of Joram et al. \cite{Jor95}, where the $S_{11}$ and $S_{31}$ 
phases were
found to be significantly smaller by about 1 degree, i.e. by 15-30 \%.
The main difference between the SE fits and the SAID or KH80
predictions shows up in the $P_{11}$ phases where we find a significant
shift to values lower by a quarter of a degree, corresponding to
a change in the phase by more than 30 \%. Of course, final
conclusions will have to await a full phase shift analysis.

\begin{figure}
\includegraphics[width=\columnwidth]{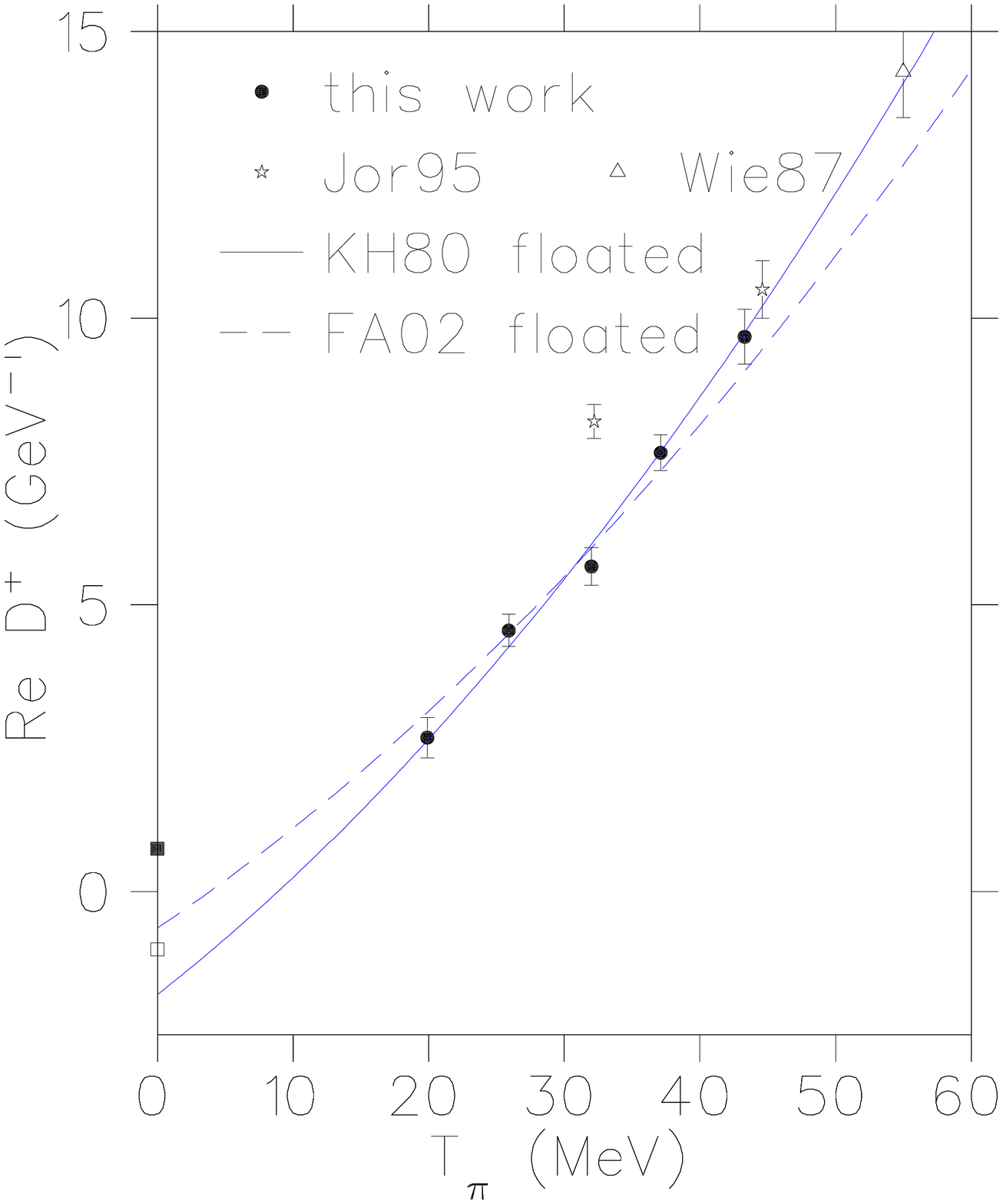}
\caption{Energy dependence of Re$D^+$.  The error bars reflect statistical
uncertainty of the data, the phase shift results have been renormalized to 
overlap the data. The original KH80 (FA02) endpoints are shown as open 
(closed) rectangles.}
\label{redplus}
\end{figure}

In analyses combining the cross sections for  $\pi^+p$ and $\pi^-p$
scattering, these data
were used to directly determine the real part of the isospin even forward 
scattering amplitude, $Re D^+$, at t = 0 as a function of incident pion 
energy, 
as was done for example in the first method of Joram et al. \cite{Jor95}. 
The scattering amplitude at threshold was determined by fitting 
the corresponding
predicted curves from phase shift analyses to the data, as shown 
in fig. \ref{redplus} compared with the 
predictions of KH80 and earlier data \cite{Jor95}, \cite{Wie87}. 
The $a^+_{0^+}$ determined by fitting the KH80 
results to the data is $(-0.126 \pm 0.010) GeV^{-1}$, shifted by -0.053
$GeV^{-1}$ from the KH80 result (solid line). The corresponding value using the functional 
form of FA02 (dashed line) is $(-0.044 \pm 0.010) GeV^{-1}$, shifted by -0.093
$GeV^{-1}$. Although these shifted values of the scattering length 
correspond to a $\pi$N-sigma term at the low end of the range currently being
discussed, it is very important to recognize that 
such extracted physics quantities are best determined through a full PSA, also
making use of the complementary data available at energies above 
those of this work. In the low energy region
the present experiment yields an extensive set of $\pi p$ cross sections that
almost triples the amount of pion proton cross sections. 
Together with the recent results on analyzing powers \cite{Rudi, Jeff}  
and pionic hydrogen \cite{Schroeder}, it provides a
much expanded data base for the determination of the phase shift solutions,
extraction of the scattering lengths and $\pi$N sigma term. 

\begin{acknowledgments}
We gratefully acknowledge support from the German ministry for education and 
research (BMBF grant no. 06TU987 and 06TU201), the Deutsche 
Forschungsgemeinschaft (DFG: Europ. Graduiertenkolleg 683, 
Heisenbergprogramm), TRIUMF, the Natural Sciences and Engineering Research
Council of Canada, the Instituto Nazionale di Fisica Nucleare (INFN), the
Australian Research Council and the California State University Sacramento 
Foundation.

\end{acknowledgments}

\end{document}